# Direct Patterning of Boron-doped Amorphous Carbon Using Focused Ion Beam-assisted Chemical Vapor Deposition

Ryo Matsumoto[a,b], El Hadi S. Sadki[c], Hiromi Tanaka[d], Sayaka Yamamoto[a,b],
Shintaro Adachi[a], Hiroyuki Takeya[a], and Yoshihiko Takano[a,b]

[a]National Institute for Materials Science, 1-2-1 Sengen, Tsukuba, Ibaraki 305-0047, Japan
[b]Graduate School of Pure and Applied Sciences, University of Tsukuba, 1-1-1 Tennodai, Tsukuba, Ibaraki 305-8577, Japan
[c]Physics Department, College of Science, United Arab Emirates University, Al Ain UAE
[d]National Institute of Technology, Yonago College, 4448 Hikona, Yonago, Tottori 683-8502, Japan

**Abstract**

The deposition of boron-doped amorphous carbon thin films on $SiO_2$ substrate was achieved via a focused ion beam-assisted chemical vapor deposition of triphenyl borane ($C_{18}H_{15}B$) and triphenyl borate ($C_{18}H_{15}BO_3$). The existence of boron in the deposited film from triphenyl borane, with a precursor temperature of 90 °C, was confirmed by a core level X-ray photoelectron spectroscopy analysis. The film exhibited a semiconducting behavior with a band gap of 285 meV. Although the band gap was decreased to 197 meV after an annealing process, the film was still semiconductor. Additionally, a drastic reduction of the resistance on the deposited film by applying pressures was observed from an in-situ electrical transport measurements using a diamond anvil cell.

## 1. Introduction

Carbon materials, which exhibit three allotropes of graphite, diamond and fullerenes, attract considerable attention in various research fields. Boron is one of the most effective dopants into carbon materials to induce hole carriers. For example, a remarkable electrical resistivity reduction of an individual multiwalled carbon nanotube was observed by boron-doping [1]. Such boron-doped carbon materials are especially interesting from the viewpoint of superconductivity. According to the McMillan relation, superconducting transition temperature $T_c$ is proportional to the Debye temperature [2]. Boron-doped diamond, which exhibits a high Debye temperature, is well established with observed $T_c$ ranging from 4 to 11 K with increasing boron concentration in substitutional lattice sites of diamond [3-7]. Recently, superconductivity with $T_c$ of 36 K was reported in heavily boron-doped amorphous quenched carbon for 17.0 ± 1.0 atom % boron concentration [8]. The $T_c$ in the boron-doped amorphous quenched carbon could be enhanced up to 55 K with further boron-concentration of ~67% [9]. The boron-doping technique is focused on not only the explorations for superconducting materials but also wide research fields, for example, a remarkably thermal sensitive property was reported on the resistivity in the boron-doped amorphous carbon [10]. It can be applicable for a thermometer because the boron-doped amorphous carbon exhibits higher temperature coefficient of resistance α, which parameter is used to define the thermometer sensibility, than that of regular NbN thin films, a common thermometer used for sensitive thermal measurements [11,12].

In this study, we proposed novel synthesis and direct patterning technique for the boron-doped amorphous carbon by using focused ion beam (FIB). One of the most common applications of FIB is the deposition of metallic or insulating films from the induced decomposition of a chemical precursor over a substrate by the ion beam, so-called FIB-assisted chemical vapor deposition (FIB-CVD). The main advantages of FIB-CVD are the deposition of the desired patterns of films without the need of a template, such as a mask or resist. The FIB-CVD is also useful for fabrication of superconducting microcircuit. The deposited tungsten thin film from a precursor of $W(CO)_6$ using FIB-CVD shows $T_c$ around 6 K [13]. This is an applicable technique for making superconducting devices, such as a superconducting quantum interference device (SQUID) magnetometer [14].

In the FIB-CVD process, carbon is naturally present in the deposited films caused by activated carbon from organic precursors [13]. The superconducting tungsten films also include 40% of carbon atoms, which is a key factor of higher $T_c$ than that of bulk tungsten of around 0.01 K. Moreover, it is reported that the deposited carbon films from the aromatic hydrocarbon precursor by FIB-CVD are composed of amorphous carbon [15]. If a film can be obtained from organic precursors, including boron by the FIB-CVD process, the resulting product would possibly contain amorphous carbon, namely a boron-doped amorphous carbon. It is a novel idea to realize direct patterning of boron-doped amorphous carbon via FIB-CVD process.

## 2. Experimental procedures

The FIB-CVD experiments are carried out using a SMI9800SE (Hitachi High-Technologies) FIB machine, equipped with a triphenyl borane ($C_{18}H_{15}B$) and triphenyl borate ($C_{18}H_{15}BO_3$) precursors in the reservoir units with a heater connected to the vacuum chamber by means of a nozzle. A silicon substrate, covered by SiO$_2$ layer with gold microelectrodes for electrical transport measurements was used in the deposition process. The operating Ga$^+$ beam energy is 30 keV. The deposition FIB current is set to 292 pA, with an aperture size of 150 μm. The precursor temperature was systematically changed from room temperature to the melting temperature of the deposition sources. During the deposition, the chamber pressure is kept at 10$^{-3}$-10$^{-4}$ Torr. In all the experiments, a FIB dose of 4.5×10$^{16}$ C/cm$^2$ is used for calibration. In order to determine the bonding state of deposited carbon films, a Raman spectroscopy analysis was carried out by using an inVia Raman Microscope (RENISHAW) with a 532 nm wavelength laser as the excitation source. Chemical composition of the obtained thin films was estimated by X-ray photoelectron spectroscopy (XPS) analysis using AXIS-ULTRA DLD (Shimadzu/Kratos) with AlK$\alpha$ X-ray radiation ($hv$ = 1486.6 eV), operating under a pressure of the order of 10$^{-9}$ Torr. The analyzed area was approximately 1×1 mm$^2$. The background signals were subtracted by active Shirley method using COMPRO software [16]. The electrical transport properties were measured by a standard four-terminal method using a physical property measurement system (Quantum Design: PPMS) under ambient and high-pressure conditions.

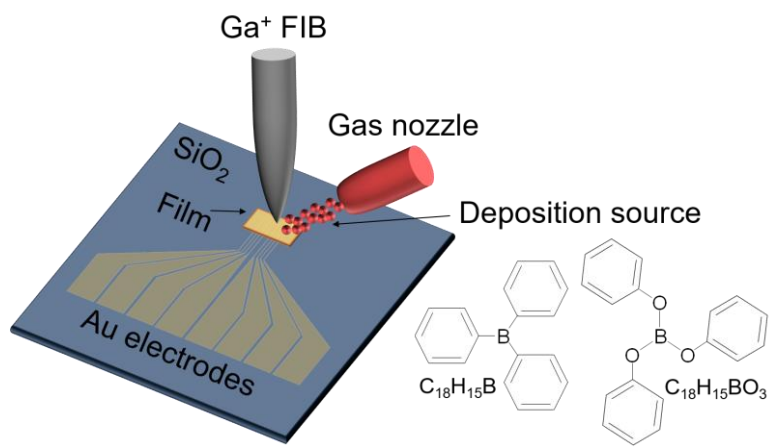

**Figure 1. Schematic image of the deposition process for boron-doped amorphous carbon film by the FIB-CVD from triphenyl borane (C$_{18}$H$_{15}$B) and triphenyl borate (C$_{18}$H$_{15}$BO$_3$) precursors.**

**3. Results and discussions**

Figure 2 shows optical microscope images of the deposited films. Deposition sources and precursor temperatures are (a) triphenyl borate and room temperature, (b) triphenyl borate and 50 °C, (c) triphenyl borane and room temperature, and (d) triphenyl borane and 90 °C, respectively. For all deposition sources and precursor temperatures, films were successfully deposited on the SiO$_2$ substrate with gold electrodes. The films from triphenyl borate showed blue color and the film from triphenyl boron of room temperature showed red color. The film from triphenyl borane of 90 °C only showed black color. Hereinafter, the film from triphenyl borane of 90 °C is called "black film" in this paper.

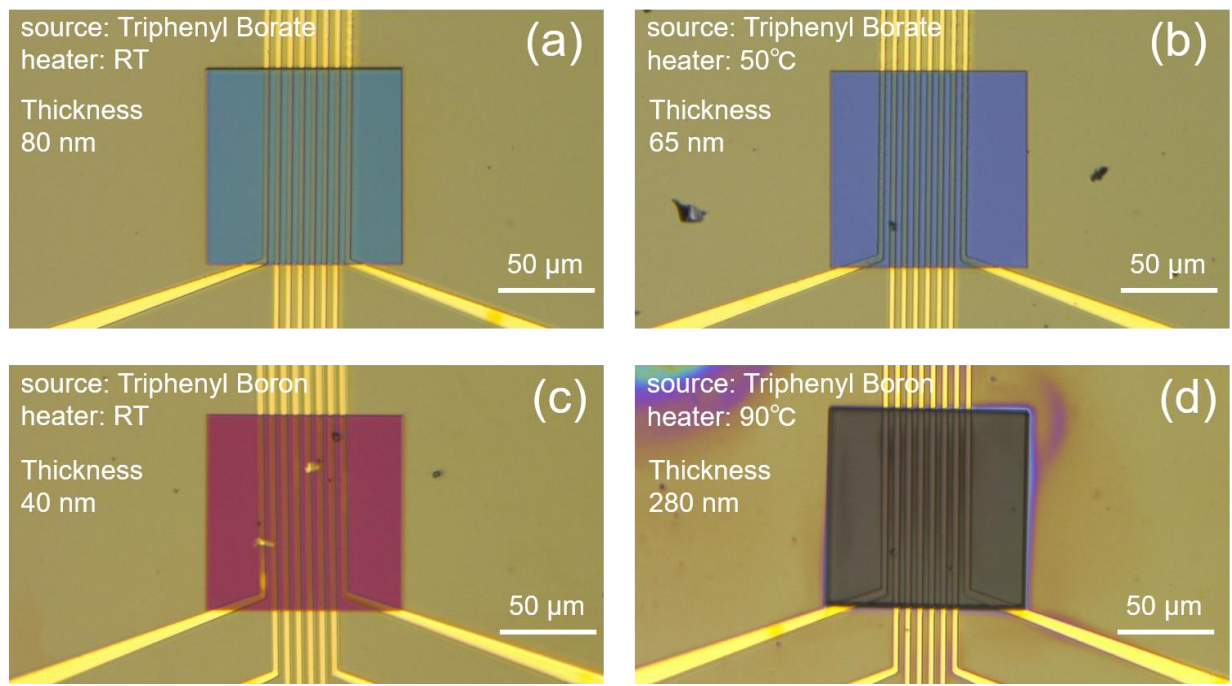

**Figure 2.** Optical microscope images of the deposited films. Deposition sources and precursor temperatures are (a) triphenyl borate and room temperature, (b) triphenyl borate and 50 °C, (c) triphenyl borane and room temperature, and (d) triphenyl borane and 90 °C, respectively.

Figure 3 (a) shows an XPS survey spectrum of the black film. All peaks were labeled by C, Ga, Si, and Au, originated from amorphous carbon, contamination from gallium ion beam, substrate, and electrodes, respectively. Figure 3 (b) shows high-resolution core-level XPS spectra around B 1s orbital of the deposited films from (i) triphenyl borate of room temperature, (ii) triphenyl borate of 50 °C, (iii) triphenyl borane of room temperature, and (iiii) triphenyl borane of 90 °C (black film). According to the spectra except for that from the black film, clear B 1s peaks were not detected, meaning that the boron-concentrations in them were less than the detection limit of XPS analysis. In contrast, the spectrum from the black film clearly showed an individual peak around 192 eV, which is the known energy for the boron [17], indicating the presence of boron atoms in the black film. In general, the detection limit of the XPS analysis is around 1 atomic % [18], suggesting the boron-concentration in the black film is above 1 atomic %. Here we note that the XPS analysis is surface sensitive with the depth resolution of few nm. The appearance of the boron peak is not depending on the film thickness and indicates an intrinsically higher concentration of boron in the black film.

To identify the ratios of $sp^3$- and $sp^2$-hybridized carbons, the peak separation was carried out for the core-level XPS spectrum of C 1s orbital from the black film as shown in Fig. 3 (c). The peaks were labeled C1, C2, and C3 peaks according to the method established by Honda et al [19]. The main peak of the maximum intensity after peak separation was assigned to the carbon with $sp^2$-hybrids in the bulk of amorphous carbon (C1 peak). The peak attributable to the carbon with $sp^3$-hybrids in the bulk of amorphous carbon (C2 peak) was observed at high binding energy side of +0.7 eV. Herein, a successful deposition the boron-doped amorphous carbon film by using FIB-CVD process was confirmed. Much detailed analysis, for example, quantitation of boron amount is required as a further study using a secondary ion mass spectrometry (SIMS).

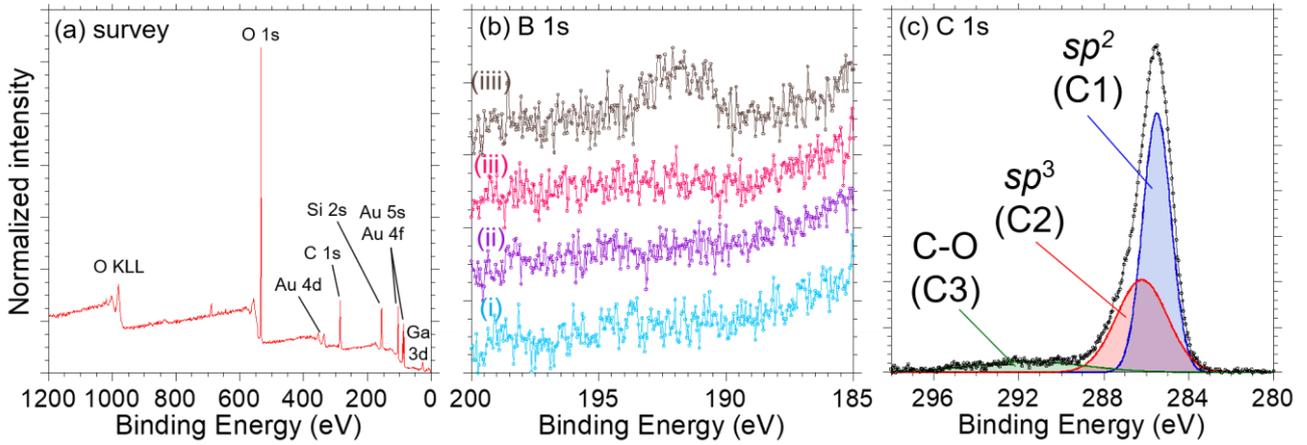

**Figure 3.** (a) Survey spectrum of the black film. (b) Core-level XPS spectra around B 1s orbital in the deposited films from (i) triphenyl borate of room temperature, (ii) triphenyl borate of 50 °C, (iii) triphenyl borane of room temperature, and (iiii) triphenyl borane of 90 °C (black film). (c) Core-level XPS spectrum of C 1s orbital from the black film.

Figure 4 (a) shows the temperature dependence of the resistance for the black film. The labeled band gap was derived from the Arrhenius plot of $R=R_0\times\exp(E_a/k_BT)$, where $R_0$ is the residual resistance, $E_a$ is the activation energy, $k_B$ is the Boltzmann constant and $T$ is the temperature, as shown in the inset. The sample exhibited a semiconducting behavior with a narrow band gap of $E_g = 2E_a \sim 280$ meV near room temperature. It can be considered that the semiconducting behavior is caused by low boron concentration in the amorphous carbon film. To decrease the band gap of the film, we performed $O_2$ annealing at 500 °C for 1 hour. It is expected that the boron-concentration in the carbon film could be increase by the annealing process because of the evaporation of carbon as $CO_2$. The transport properties of the annealed film are shown in fig. 4 (b). The resistance and band gap were dramatically decreased after the annealing in accordance with our scenario. Although the temperature dependence of resistance is still semiconducting, it would be possible to metallize the property by tuning the annealing conditions.

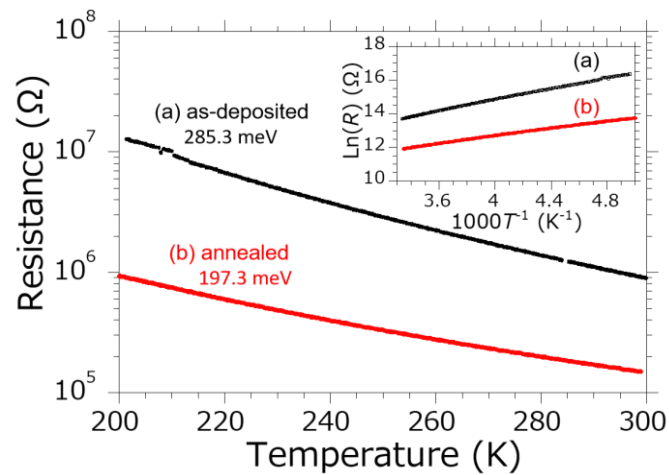

**Figure 4.** Temperature dependence of resistance in (a) as-deposited film and (b) annealed film from triphenyl boron at 90 °C. The inset shows Arrhenius plot of the temperature dependence of resistance.

Figure 5 shows Raman spectrum of the annealed black film. According to the established peak separation method in boron-doped amorphous Q-carbon [8], the spectrum can be deconvoluted into three Raman-active vibrational modes. The first peak around 1150 cm$^{-1}$ corresponds to sp$^2$ bonding, whereas the second peas around 1370 cm$^{-1}$ and third peak around 1590 cm$^{-1}$ represent sp$^3$-bonded carbon and graphitic phase (G peak), respectively. These peak components indicate that the annealed film is amorphous carbon.

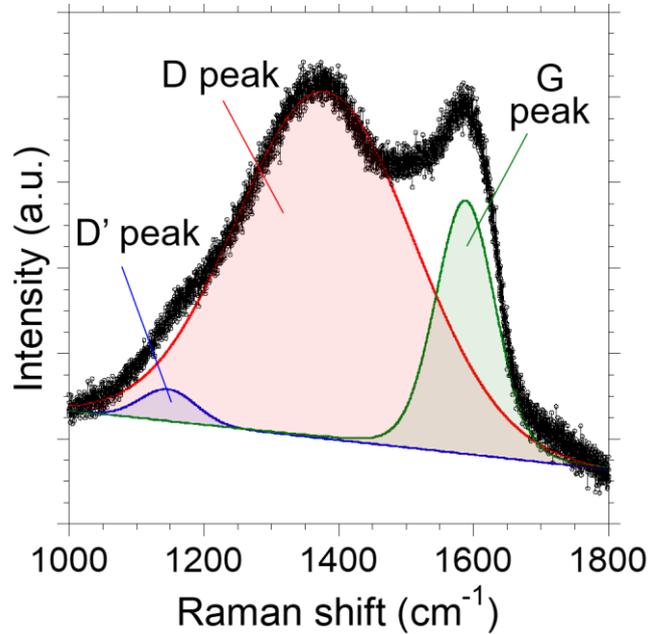

**Figure 5. Raman spectrum of the annealed black film and deconvoluted peaks.**

As one of the ways to metallize the carbon film, we examined the application of high-pressure to improve the connectivity in the film. We used a diamond anvil cell (DAC) with boron-doped diamond electrodes [20,21] to measure the transport property under high pressure. The inset of fig. 6 is an optical microscope image of the film from triphenyl borate of room temperature on the bottom diamond anvil with diamond electrodes. The stainless-steel gasket and pressure-transmitting medium of boron-nitride powder were used in the high-pressure generation. The pressure values were determined from a pressure dependence in the wavelength of fluorescence from ruby powder in the sample space [22].

Figure 6 shows temperature dependence of resistance under various pressures in the film up to 20.7 GPa. Although the behavior is still semiconducting even under 20 GPa, the film resistance drastically decreased with increase of the applied pressure as expected. To realize the metallization and superconducting transition, it is necessary to investigate suitable deposition sources, and the deposition and annealing conditions. If the superconductivity appears in the film by FIB-CVD, it is the innovative method to draw directly the high-$T_c$ superconducting pattern on desired substrate without any masking process, such as a lithography.

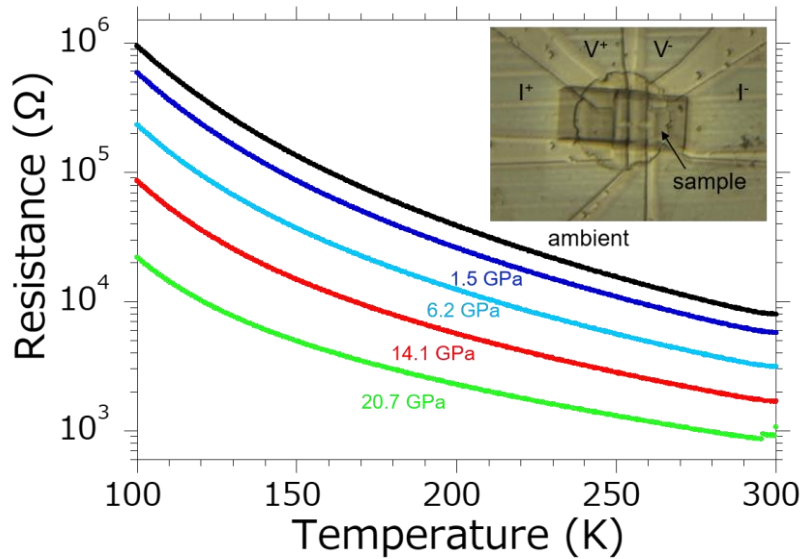

**Figure 6.** Temperature dependence of resistance under high pressures up to 20.7 GPa in the film from triphenyl borate of room temperature. The inset is an optical microscope image of the film on the bottom diamond anvil with diamond electrodes.

## 5. Conclusion

The unique pathway to achieve boron-doped amorphous carbon was considered via FIB-assisted CVD method. The core level XPS analysis revealed that the only black film from the deposition source of triphenyl borane with the precursor temperature of 90°C contained boron. Although the black film showed relatively low resistance after annealing process, superconductivity was not observed. It is necessary to adjust the deposition and annealing conditions to reduce the resistance. On the other hand, the high-pressure application decreased the resistance of the deposited film dramatically. The combination of the annealing and high-pressure application would adjust the functionalities of the films more effectively.

**Acknowledgment**

This work was partly supported by JST CREST Grant No. JPMJCR16Q6, JST-Mirai Program Grant Number JPMJMI17A2, JSPS KAKENHI Grant Number JP17J05926 and 19H02177. A part of the fabrication process of diamond electrodes was supported by NIMS Nanofabrication Platform in Nanotechnology Platform Project sponsored by the Ministry of Education, Culture, Sports, Science and Technology (MEXT), Japan.